# Resonant frequency analysis of dental implants

D. Rittel[1*], A. Dorogoy[1], G. Haïat[2] and K. Shemtov-Yona[1]

[1] Faculty of Mechanical Engineering, Technion, 32000, Haifa, Israel

[2] CNRS, Laboratoire Modélisation et Simulation Multi-échelle, UMR CNRS 8202, 94010 Créteil Cedex, France

**Abstract**

Dental implant stability influences the decision on the determination of the duration between implant insertion and loading, This work investigates the resonant frequency analysis by means of a numerical model.

The investigation is done numerically through the determination of the eigenfrequencies and performing a steady state response analyses using a commercial finite element package. A peri-implant interface, of *simultaneously* varying stiffness (density) and layer thickness is introduced in the numerical 3D model in order to probe the sensitivity of the eigenfrequencies and steady state response to an evolving weakened layer, in an attempt to identify the bone reconstruction around the implant.

For the first two modes, the resonant frequency is somewhat insensitive to the healing process, unless the weakened layer is rather large and compliant, like in the very early stages of the implantation. A "Normalized Healing Factor" is devised in the spirit of the Implant Stability Quotient, which can identify the healing process especially at the early stages after implantation.

The sensitivity of the RFA to changes of mechanical properties of periprosthetic bone tissue seems relatively weak.

Another indicator considering the amplitude as well as the resonance frequency might be more adapted to bone healing estimations. However, these results need to be verified experimentally as well as clinically.

## 1. INTRODUCTION

Dental implant stability is critical for the surgical success (Haïat et al. 2014) and depends on the quantity and biomechanical quality of the peri-implant bone tissue (Franchi et al. 2007). Two kinds of implant stability can be distinguished. Primary stability occurs just after implant surgery, when the implant in inserted within bone tissue. Dental implant primary (immediate) stability should be sufficiently important in order to restrain micromotion at the bone-implant interface, but should not be too high because it might cause bone necrosis (Mathieu et al. 2014). Secondary (final) stability is obtained through osseointegration phenomena, of a complex multi-time and multiscale nature, which strongly depends on the implant primary stability.

Dental implant stability is still difficult to assess clinically because it depends on the implant properties, the patient bone quality and the surgical protocol. Dental implant stability influences the decision on the determination of the duration between implant insertion and loading, which may vary from 0 up to 6 months (Raghavendra, Sangeetha; Wood, Marjorie C.; Taylor 2005). A compromise should be found in a patient-related manner based on early implant loading. The rationale is to stimulate osseointegration phenomena, and also apply late implant loading in order to avoid degradation of the consolidating bone-implant interface in early postsurgical stages (Serra et al. 2008). Shortening the time to implant loading has thus become a challenge in recent implant developments to both minimize the time of social disfigurement and avoid gum loss. Therefore, accurate measurements of implant biomechanical stability are of interest since they could be used to improve the surgical strategy by adapting the choice of the healing period in a patient-specific manner.

Assessing the implant stability is a difficult multiscale problem because of the complex heterogeneous nature of periprosthetic bone tissue and the involved bone remodeling phenomena (Wolff 1986; Frost 2003). Different approaches have been used to assess implant stability *in vivo*. So far, most surgeons still rely on their proprioception because it remains difficult to monitor bone healing *in vivo* (Serra et al. 2008). Accurate quantitative methods capable of assessing implant stability are required to guide the surgeons and eventually reduce the risk of implant failure.

Magnetic resonance imaging (Hecht et al. 2011) as well as X-ray based (Albrektsson et al. 1988) techniques are of limited interest because of diffraction phenomena occurring at the bone-implant interface due to the presence of metal. Therefore, alternative biomechanical methods have been developed, their main advantage being the absence of ionizing radiation, inexpensiveness, portability and noninvasiveness. The measurement of the insertion torque to assess dental implant primary stability has often been evoked, but this approach remains limited (Bayarchimeg et al. 2013) because, while the result is globally related to the properties of the bone-implant interface, it does not directly reveal implant micromotions and cannot be used for secondary stability assessment. The Periotest (Bensheim, Germany) is a percussion test method (SCHULTE & W 1983; Van Scotter & Wilson 1991). Its sensitivity to striking height and handpiece angulation complicates the clinical examination (Meredith, Neil; Friberg, Bertil; Sennerby, Lars; Aparicio 1998) and limits the reproducibility of the measurements.

Another method consists of using quantitative ultrasound (QUS) (De Almeida et al. 2007) to investigate the properties of the bone-implant interface. The principle of the measurement relies on the dependence of ultrasonic propagation within the implant on the boundary conditions prescribed by the biomechanical properties of the bone-implant interface (Mathieu et al. 2011b) (Vayron et al. 2013) (Vayron et al. 2018b) (Vayron et al. 2014a) (Vayron et al. 2014b) (Vayron et al. 2018a) (Mathieu et al. 2011a) (Vayron et al. 2015, 2016; Hériveaux et al. 2018).

The most commonly used biomechanical technique is the resonant frequency analysis (RFA) (Valderrama et al. 2007), which consists of measuring the first bending resonance frequency of a small rod attached to the implant (Meredith et al. 1996). The RFA technique allows to assess the implant anchorage depth into bone (Meredith et al. 1997), marginal bone level (Friberg et al. 2003) and the stiffness of the bone-implant structure (Ersanli et al. 2005; Pattijn et al. 2007).

Different clinical studies have also been carried out in order to determine the threshold in terms of resonance frequency above which i) an implant can be considered to have a good primary stability and ii) an implant can safely be loaded (Baltayan et al. 2016). In those measurements, the system provides an "Implant Stability Quotient (ISQ)" whose value is supposed to reflect the quality of the implant stability in the jawbone. While the ISQ has been widely reported in the literature, its exact definition remains obscure as it is not defined in any scientific publication and remains a proprietary information of the system manufacturer. Yet, the sensitivity of ISQ to variation of bone properties around the implant has not been investigated in detail, which is difficult to achieve using experimental approaches only. Therefore, numerical simulations, using e.g. finite element modeling, are useful to estimate the overall vibration pattern (eigenfrequencies and mode shapes) of the bone-implant system, including cases where the peri-implant properties may vary as a result of the healing and osseointegration process vary. Incidentally, such simulations could be used to determine an ISQ for each simulated case.

Finite element numerical simulation tools have already been used to show that the orientation and fixation of the transducer have an important effect on ISQ values [31] obtained with the older Osstell version with an L-shaped, wired transducer. Perez et al. (Pérez et al. 2008) modeled the resonant response of a bone implant system in which the interface is subjected to an evolution that reflects bone healing. In this model, the interface has no real physical thickness, but the results compare favorably with clinical results, while the reported frequencies are in the 2-4 KHz range, with a marked sensitivity of the interfacial state. Harirforoush et al. (Harirforoush et al. 2014) emphasized the influence of the implant angulation angle on the resonant frequency. By contrast, the aforementioned authors reported a significant influence of the relative contact area of the implant with the cortical and trabecular bone components, which was varied through the angulation process. In their model, the authors did not explicitly consider a bone-implant interface of any specific kind. Li et al. (Li et al. 2011) investigated the effect of bone remodeling using a criterion formulated in terms of strain energy density. In this work, bone remodeling was not restricted to a specific interface of an evolving thickness, but rather estimated (updated) from the bulk stress distribution of the strain energy density of the modelled bone-implant system that is subjected to specific mechanical constraints. In that work, both the bone stiffness and density evolved concurrently. More recently, Zanetti et al (Zanetti et al. 2018) investigated the influence of implant design on the changes of the resonance frequency of bone-implant system during osseointegration using modal analysis. Their thorough analysis of different implant shapes also considered variations of the peri-implant bone stiffness in order to mimic bone maturation. The authors reported that the first two resonant frequencies are weakly sensitive to the degree of bone maturation beyond roughly 20% for all the considered implant models. Such conclusions raise fundamental questions about the capability of the RFA method to discriminate bone healing. However the authors employed modal analysis, and the determination of the implant micromotions, and additional steady-state dynamic analysis would have provided further understanding on the implant behavior. Likewise, they did not provide a quantitative "figure of merit" for each implant in the spirit of the above-mentioned ISQ.

In this work we consider a single generic implant geometry (similar to that modelled by Hariforoush et al. (Harirforoush et al. 2014)) fully anchored in a jawbone section, in which a peri-implant bone layer is assigned various stiffness values and width. We systematically characterize the vibration modes and resonant frequencies of the model for various kinds of peri-implant degraded bone layers in order to characterize their influence on the resonant frequencies. Moreover, we also carry out a dynamic steady state analysis for the above-studied cases, in order to evaluate the implant micromotions and devise a "figure of merit", the "Normalized Healing Factor (NHF)" that helps defining the stability of a dental implant.

## 2. MATERIALS AND METHODS

A section of a mandible human bone, with a flush inserted metallic dental implant connected to a peg were modeled using the commercial finite element (FE) package Abaqus (Simulia 2014a). A full 3D modal analysis was first carried out, followed by a steady-state dynamic modal analysis.

The first analysis determines the resonant frequencies and corresponding modes of the implant. The results of the frequency extraction step are obtained by the Lanczos eigensolver (Simulia 2014b). The eigenvalue problem for natural modes of small vibration of a finite element model in a classical matrix notation is given by (Simulia 2014b) as:

$$\left(\mu^2[M] + \mu[C] + [K]\right)\phi^N = 0 \qquad (1)$$

where [M] is the mass matrix, which is symmetric and positive definite in the problems of interest here; [C] is the damping matrix; [K] is the stiffness matrix, which may include large-displacement effects, such as "stress stiffening" (initial stress terms), and, therefore, may not be positive definite or symmetric; μ is the eigenvalue; and is the eigenvector—the mode of vibration.

The second analysis is a steady-state dynamic modal (SSDM) analysis which predicts the linear response of the structure subjected to continuous harmonic excitation (Simulia 2014b). It uses the set of eigenmodes extracted in the previous eigenfrequency step to calculate the steady-state solution as a function of the frequency of the applied excitation. The analysis is done as a frequency sweep by applying the loading at a series of different frequencies and recording the response. The software conducts this frequency sweep. The frequency range which is used here is : 4000 Hz < f < 12000 Hz. The load which is applied is shown in Fig. 2a: $P_x = 1$ N, $P_z = 0$ N. This direction of the loads corresponds to the first eigenmode shown in Fig. 2a. These type of load is also applied during a typical resonant frequency test.

## 2.1 Finite Element Model

### 2.1.1 Parts and Assembly

The assembly consists of three parts, shown in Fig. 1a-c: i) the mandible bone, ii) the dental implant and iii) the peg.

The mandible bone was created by extruding a typical cross section of the mandible at the molar region by 20 mm along the Z axis (Fig. 1a). The cross section consists of a cortical bone shell of roughly ~2 mm thickness, which surrounds the internal trabecular bone.

A MIS Seven dental implant was inserted in the middle of the bone section, similarly in the surgical protocol. The implant is 3.75 mm diameter near the neck, and 13.2 mm long. Five micro-rings can be found in the region of contact with the cortical bone tissue. The implant features a conical shape with threads that reduce in thickness near the apical bottom, and 3 spiral channels near the apical bottom to support the self-tapering property of the implant. A perfect geometrical fit was imposed between the implant's geometry and bone. The reason for this assumption is the lack of clarity of the procedure used by the finite element code in the case of frictional contact conditions. The implant was inserted in the Y direction. Its upper face protrudes the Y direction by 68 μm above the bone face, as shown in Figs. 2a and 2b. The bone and implant were merged into one part.

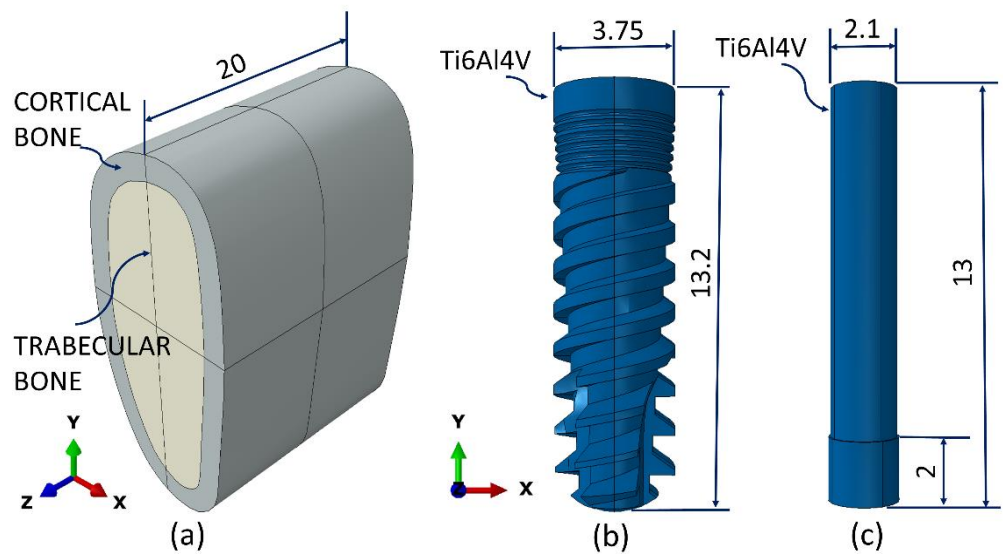

**Figure 1**: The three parts of the model: a. Mandible bone. b. Implant. c. Peg.

A peg was rigidly attached to the implant with no relative displacements between the contacting surfaces, thereby creating one part (bone+implant+peg) as shown in the exposed assembly (Fig. 2b). The peg was modelled as a simple cylinder of 2.1 mm diameter and 13 mm length. The peg geometry is similar to the commercial Osstell SmartPeg (https://www.osstell.com/product/smartpeg). The loads were applied on the top of this peg for the mode-based steady-state dynamic analysis part of this study. Loading was applied along the x direction (Px) since it corresponds to the usual test direction (see Fig. 2a).

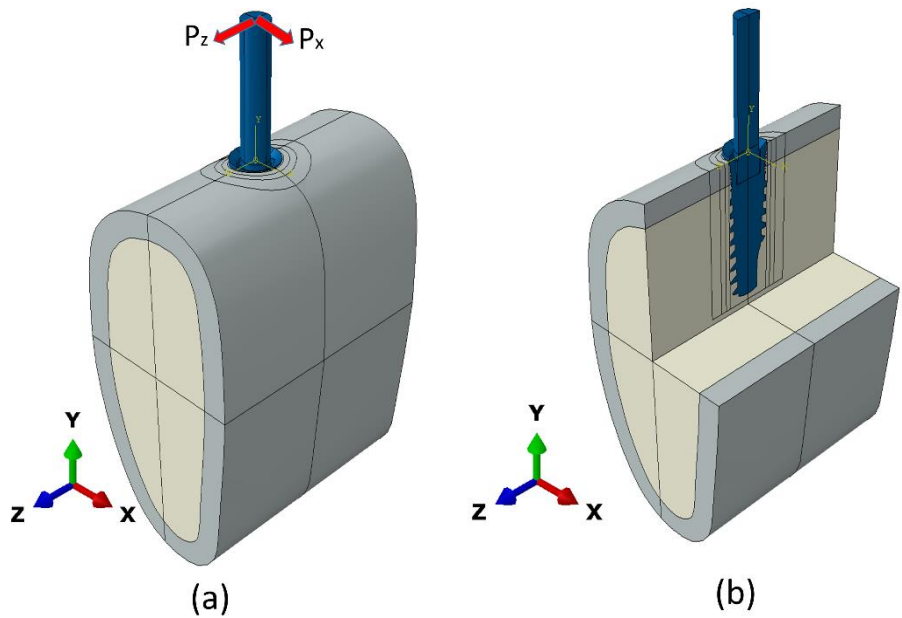

**Figure 2:** a. The assembly showing the applied loads. b. the exposed assembly showing the unified implant-peg system and the perfect geometrical fit between the implant and bone.

### 2.1.2 Material properties

For the sake of simplicity, all three parts were assigned linear elastic and homogenous material properties. For the implant and peg, isotropic mechanical properties of Ti-6Al-4V ELI(American Society for Testing and & American Society for Testing and Materials 2013) were used (see Table 1).

*Cortical* bone tissue was assumed to be isotropic with its mechanical properties according to (Guan et al. 2011) (Table 1). *Trabecular* bone tissue was also assumed to be isotropic, with mechanical properties according to (Van Staden et al. 2008; Guan et al. 2011) (Table 1).

| Material | Young's Modulus E [GPa] | Poisson's ratio $\upsilon$ | Density $\rho$ [Kg/m$^3$] |
|---|---|---|---|
| Ti-6Al-4V ELI | 113.8 | 0.33 | 4430 |
| Cortical Bone | 18 | 0.35 | 1900 |
| Cancellous bone | 0.7 | 0.34 | 1000 |

**Table 1**: Mechanical properties of the different materials used in the FE model.

### 2.1.3 Mesh and boundary conditions

The meshed model is shown in Fig. 3. The whole assembly is shown in Fig. 3a and a detail of the upper surface of the bone near the cavity is shown in Fig. 3b. This detail shows the relative position of the implant compared to the upper bone face.
The implant and peg as well as cylindrical bone region surrounding the implant were meshed with a mesh seed size of 100 µm. The dense mesh of the cavity is exposed in Fig. 3c. The exposed mesh of the implant and peg is shown in Fig. 3d. A total of 1,647,600 linear tetrahedral elements of type C3D4 were used in the model.

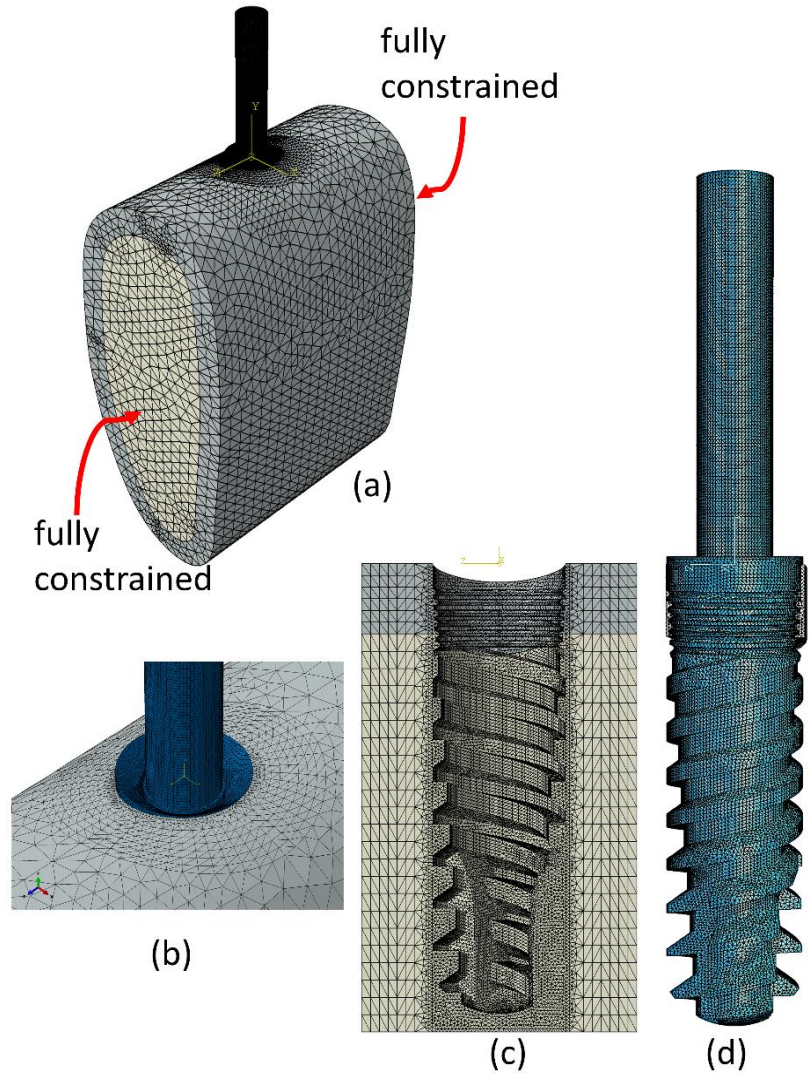

**Figure 3:** a. The meshed assembly. b. The upper face mesh near the cavity showing the relative position of the implant to the upper face of the bone. c, The exposed inner cavity bone mesh. d. The implant and peg mesh.

The two vertical sides (normal in Z direction) of the bone were constrained by application of "encastre" (fully constrained) conditions (Fig. 3a). These boundary conditions fix the assembly in space and prevent rigid body motions. For the steady state analysis, the applied load was set to Px=1N.

### 2.1.4 Parametric Studies

*Peri-implant weakened layer*
It was assumed that the peri-implant bone material properties are affected (damaged) by the insertion process, and that its stiffness ($E_i^*, i = cortical, trabecular$) decreases accordingly (Dorogoy et al. 2017). The region of interest where the material properties were modified corresponds to a cylinder, as shown in Fig 4a. Four different values of the cylinder diameter were considered, which corresponds to various values of the width *w* of the peri-implant layer: $w$ = 0.1, 0.5, 1.0 and 2.0 mm, respectively. The cross-sectional view of the regions of interest are shown in Figs. 4b-e. Note that such weakened layers might also represent the progression of osseointegration phenomena, corresponding to bone strengthening over time, and thus space.

Three different values of the stiffness $E^*$ of the peri-implant region of interest were considered: $E_i/10$, $E_i/5$ and $E_i/2$, where $E_i$ is the initial stiffness of the undamaged bone (trabecular or cortical). Both types of bones were deteriorated by the same factor. Likewise, variations of the mass density of bone tissue in the peri-implant region of interest were considered following: $\dfrac{\rho_i^*}{\rho_i} = \left(\dfrac{E_i^*}{E_i}\right)^{0.3}$ (Carter & Hayes 1977; Zanetti et al. 2018). The degraded mass density values are summarized in Table 2.

|  | Mass density $\left[\dfrac{Kg}{m^3}\right]$ | | | |
|---|---|---|---|---|
| ratio |  | $\dfrac{E_i^*}{E_i}=\dfrac{1}{2}$ | $\dfrac{E_i^*}{E_i}=\dfrac{1}{5}$ | $\dfrac{E_i^*}{E_i}=\dfrac{1}{10}$ |
| cortical | 1900 | 1508 | 1111 | 881.9 |
| trabecular | 1000 | 793.7 | 584.8 | 464.1 |

**Table 2**: The degraded mass density values used in the FE model.

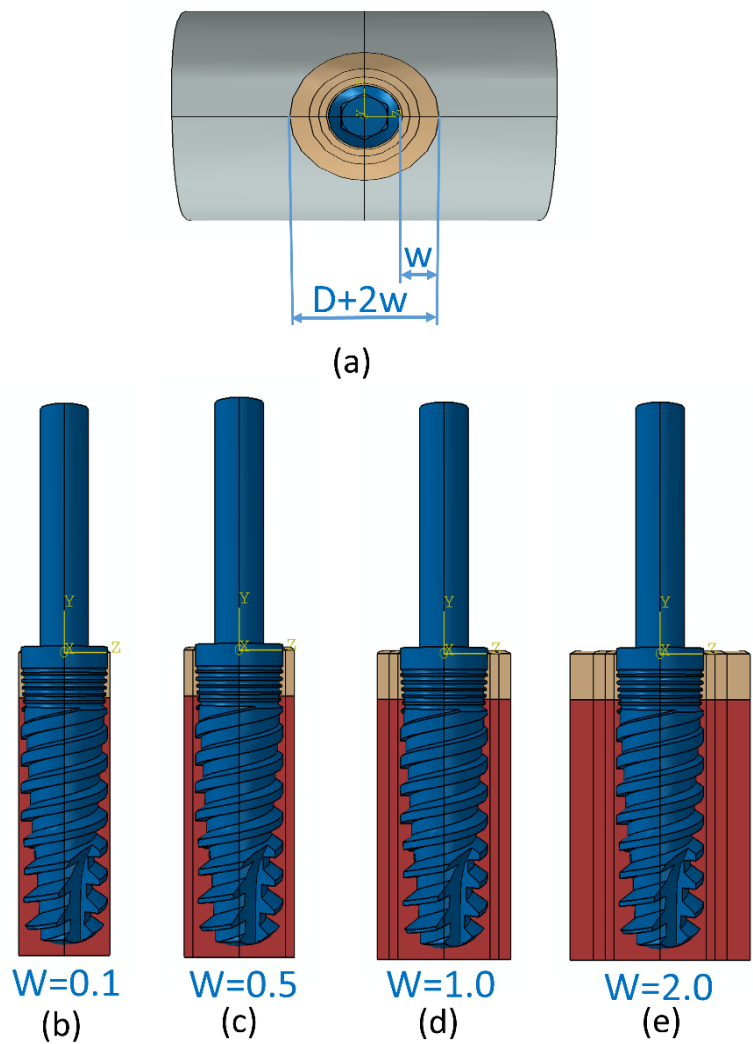

**Figure 4:** a. Top view of the assembly showing the width of the cylindrical affected zone. b. Cross-sectional view for w = 0.1mm. c. Cross-sectional view for w = 0.5 mm. d. Cross-sectional view for w = 1.0 mm. e. Cross-sectional view for w = 2 mm. Note what the case where w=0 mm corresponds to no weakened layer.

## 3. RESULTS

### 3.1 Effect of peri-implant layer stiffness, density and width.

The first 10 eigenfrequencies obtained for 5 widths of the region of interest: w = 0, 0.1, 0.5, 1.0 and 2.0 mm are fully detailed in tables A1, A2 and A3 of Appendix A. Note that w = 0 mm refers a non- weakened layer. Tables A1-A3 list the resonant frequencies for a weakened layer of $E_i^* = E_i / 10$, $E_i^* = E_i / 5$ and $E_i^* = E_i / 2$, and their corresponding densities shown in Table 2.

The first two eigenfrequencies are the most interesting since they can be monitored more easily, as higher order modes become more complex to observe. Moreover, the first two eigenfrequencies correspond to the modes actually excited by the Osstell device. These eigenfrequencies are plotted versus the peri-implant layer's width in Fig. 5. The markers represent the numerical values. The solid (respectively dashed) lines correspond to the first (respectively second) eigenfrequency. The resonant frequency values decrease as a function of *w*, and the sensitivity of the resonant frequency is higher for weaker and wider layers.

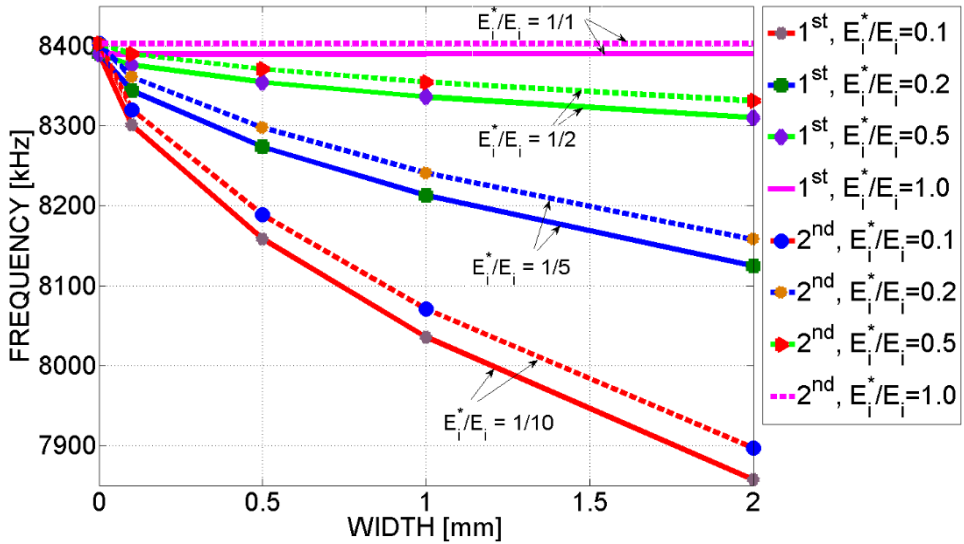

**Figure 5:** The first two eigenfrequencies as a function of the width of the peri-implant layer, 0 < *w* < 2.0 mm for Ei* = Ei / 10, Ei* = Ei / 5 ,Ei* = Ei / 2,Ei* = Ei and their corresponding densities. The solid lines represent the first eigenfrequency while the dashed represent the second.

Both the bone stiffness and the width *w* of the weakened layer affect the values of the system eigenfrequencies. The first 2 eigenfrequencies corresponding to all stiffnesses and small (<0.2 mm) thickness layers are comparable and lie below 10 kHz. When the width of the weakened layer exceeds 0.5 mm, the different stiffness values have a clear distinguishable influence. The value of the third resonant frequency is more than twice that of the first two ones for all stiffnesses. The remaining 8 eigenfrequencies (3-10) lie in the region 16836 Hz < f < 37111 Hz for all the assumed stiffnesses.

As an illustration, the 4$^{th}$, 7$^{th}$ and 10$^{th}$ eigenfrequencies as a function of the width of the weakened layer and its stiffness/density are shown in Fig. 6. The higher eigenfrequencies exhibit the same behavior as the first two ones i.e. the eigenfrequency decreases with the width of the weak layer and increases with the strength of the weak layer. The weaker and wider the affected layer, the lower the eigenfrequency.

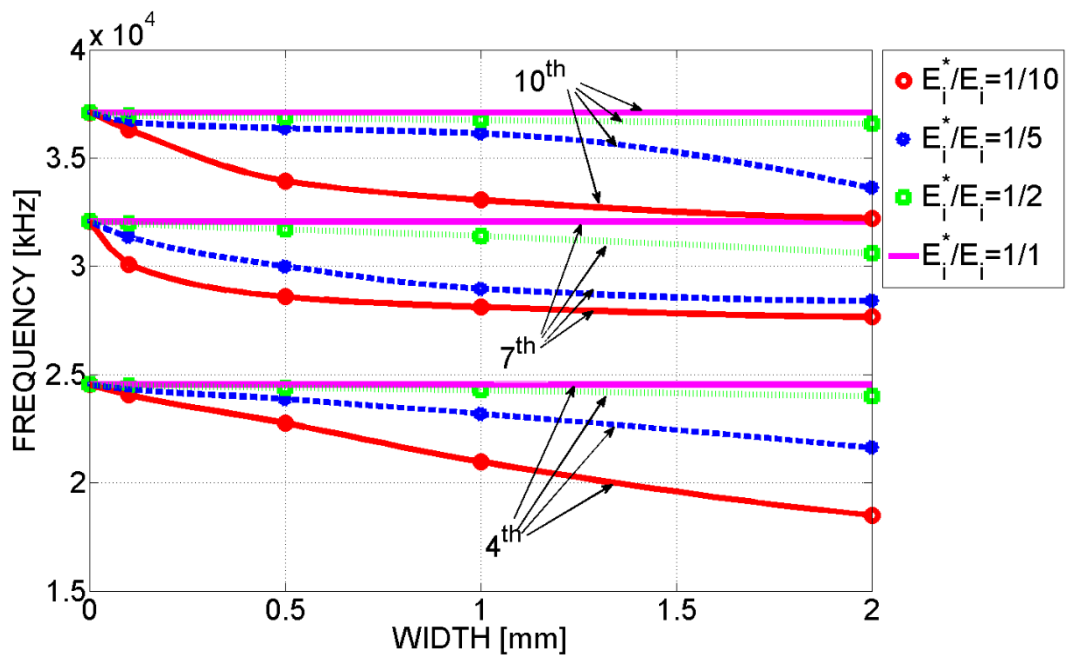

**Figure 6:** The 4$^{th}$, 7$^{th}$ and 10$^{th}$ eigenfrequencies as a function of the width of of the peri-implant layer, $0 < w < 2.0$ mm for Ei* = Ei / 10, Ei* = Ei / 5, Ei* = Ei / 2 and Ei* = Ei.

The first two eigenmodes are shown in Fig. 8. The first mode corresponds to the peg displacement in the X direction, while the second corresponds to the peg displacement in the Z direction.

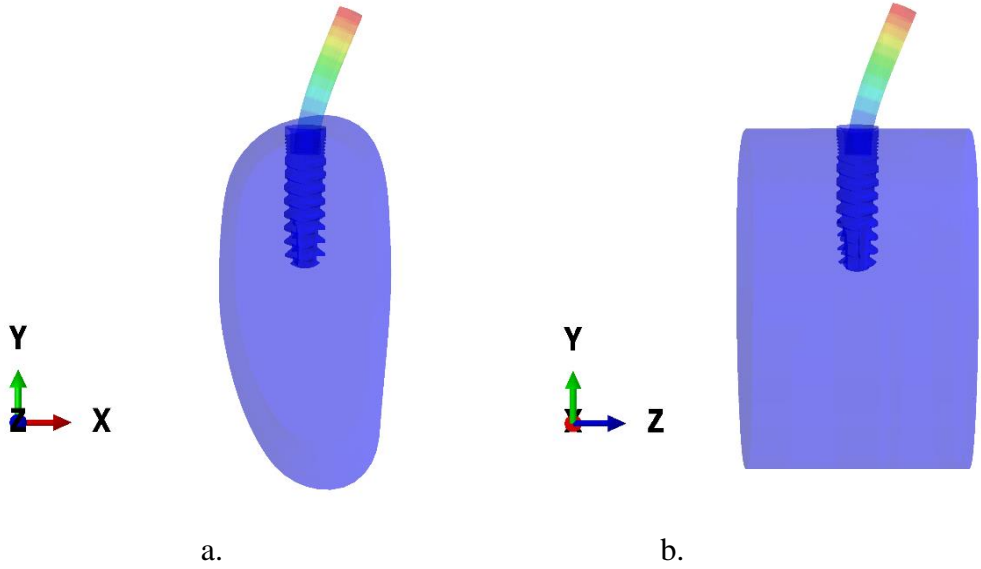

a.                  b.

**Figure 7**: a. Side view of the displacement mode which correspond to the first eigen frequency. b. The displacement mode which correspond to the second eigenfrequency.

During a healing process, the stiffness and density of the weakened layer increases and at full recovery, $E_i^* = E_i$ and $\rho_i^* = \rho_i$. At the same time, the width of the weakened layer might decrease if the healing process progresses inwards towards the implant. Figure 9 shows contour maps of the first two eigenfrequencies versus both the width and stiffness of the weakened layer. The values on the maps are linearly interpolated from the values of Tables A1-A3. The eigenfrequencies change during a healing process can be regarded as moving on the contour map from the bottom right side to the top left corner where $E_i^* = E_i$ and $w = 0$, as shown by the arrows in Figures 9a and 9b. The frequency in that top left corner is 8390 Hz (Tables A1-A3). The maximum values of the $1^{st}$ and $2^{nd}$ eigenfrequencies are 8390 Hz and 8403 Hz (Tables A1-A3). The minimum values of the $1^{st}$ and $2^{nd}$ eigenfrequencies are 7857 Hz and 7896 Hz (Table A1), respectively. The total difference is 533 Hz and 507 Hz for each mode, respectively. The maps of Fig. 9 clearly show that during most of the healing process there is only a slight change of the $1^{st}$ and $2^{nd}$ eigenfrequencies. The areas for which the eigenfrequencies are above 8320 Hz is large in comparison to the areas for which the eigenfrequencies are below 8320 Hz. The global change of the $1^{st}$ and $2^{nd}$ eigenfrequencies for areas above 8320 Hz is 70 Hz and 83 Hz, while the global change for areas beneath 8320 Hz is 443 Hz and 424 Hz, respectively. Hence, high gradients in the eigenfrequency value correspond to the beginning of the healing process where the stiffness recovers from $E_i^* / E_i = 0.1$ to $\sim E_i^* / E_i = 0.4$. The gradients are higher for initially wide weak layer. Hence, these early stages of healing should eventually be more easily monitored by the eigenfrequencies.

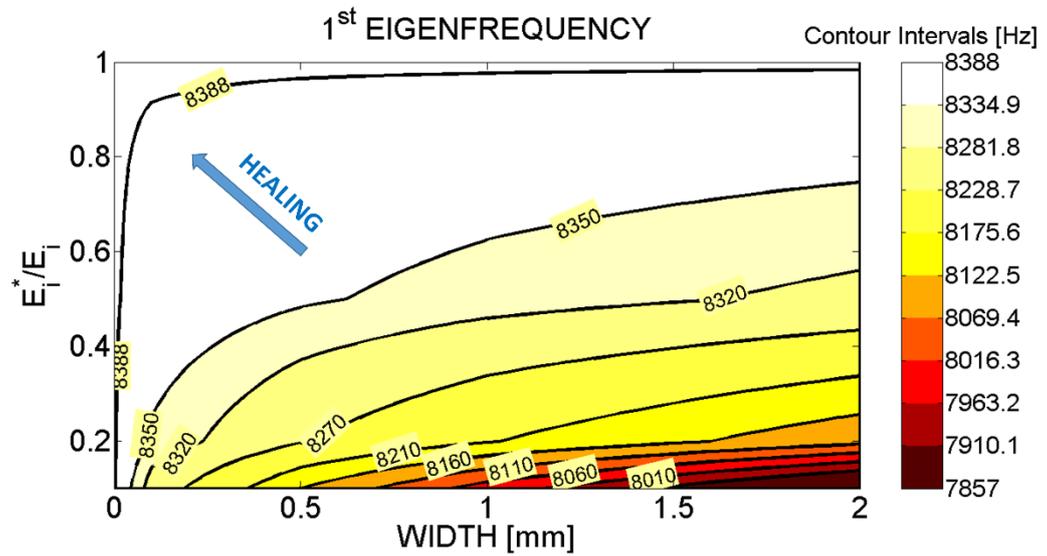

(a)

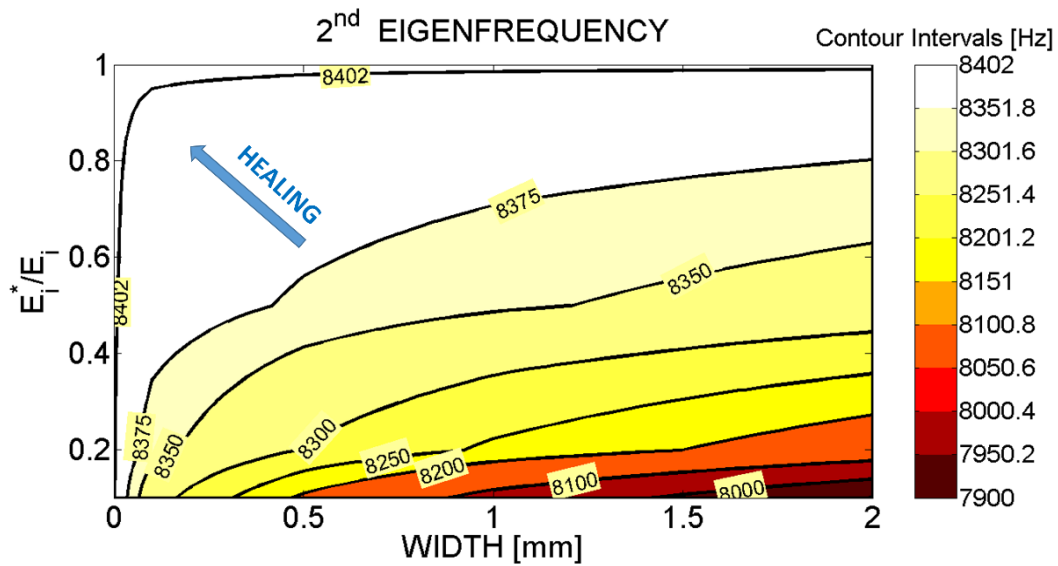

(b)

**Figure 7:** Contour maps of the first two eigenfrequencies versus the strength and width of the weakened layer. a. The 1$^{st}$ eigenfrequency. b. The 2$^{nd}$ eigenfrequency.

## 3.2 Steady state results.

The effect of the modal damping coefficient is presented first, followed by the results of the SSDM.

## 3.3 Effect of modal damping coefficient

The modal damping coefficient is an adjustable parameter of the steady state analysis. In order to assess the effect of its variation on the maximum calculated displacement, two limit cases were considered. The first case corresponds to a fully healed bone, $E_i^* = E_i$, for which w = 0 mm. The second limit case corresponds to the most degraded test bone case, setting w = 2 mm and $E_i^* = E_i/10$.

The first 20 eigenfrequencies were first determined again, followed by the SSDM analysis. Three commonly used damping factors values were: 0.05, 0.025 and 0.0125. The resulting displacement ($U_x$) is plotted in Figure 8 for a frequency range 7000 – 9000 Hz where the maximum response lies.

The damping factor has a significant effect on the amplitude, but a less important effect on the peak frequency. During a healing process the frequency increases by an amount of Δf and the amplitude decreases by an amount of ΔA, as expected from a stiffer bone-implant system. These two parameters (Δf and ΔA) are marked in Fig. 8 for the results due to a damping factor of 0.0125. The maximum range of Δf is ~500 Hz irrespective of the damping factor.

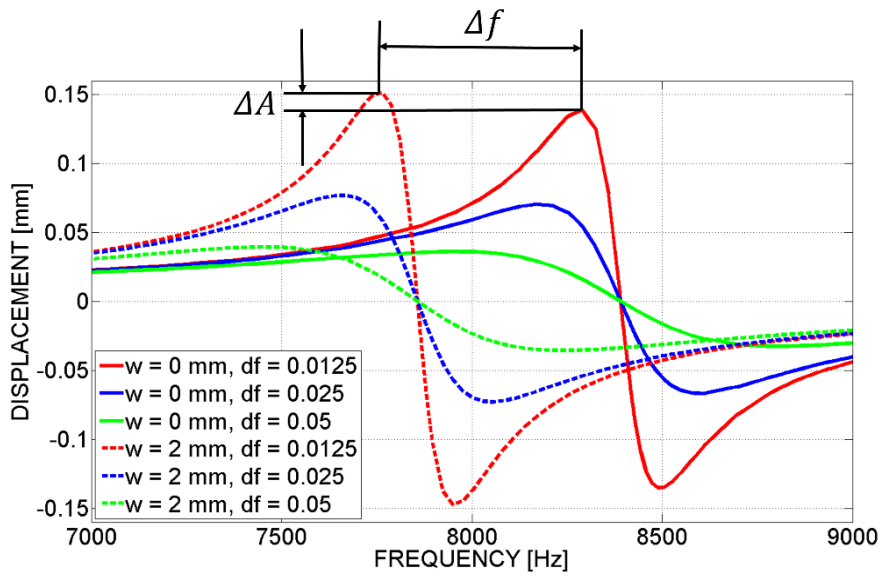

**Figure 8:** The Ux displacement at frequency range 7000 Hz < f < 9000 Hz.

The frequency and displacement results at the peak points of the curves in Fig. 8 are detailed in the 2$^{nd}$ and 3$^{d}$ rows of Table 3, while the first row lists the values of the first eigenfrequency determined in subsection 3.1. The calculated SSDM frequency values are slightly lower than the first eigenfrequency, and this difference diminishes as the damping factor is lowered.

In order to overcome this dependence on the damping factor's value, a mixed parameter p is defined:

$$p = \frac{f_m}{A_m} \quad \left[\frac{Hz}{mm}\right] \tag{2}$$

This mixed parameter combines the effect of the peak frequency to the peak amplitude. These two parameters can be measured/monitored experimentally during a healing process. The corresponding p values are detailed in the 4$^{th}$ row of Table 3.

We now normalize the p value by $p_D$ of the most degraded case (w = 2m $E_i^* = E_i/10$), defining $p^* = \frac{p}{p_D}$. The normalized $p^*$ values are presented in the 5$^{th}$ row of Table 3.

|  | w = 0 mm, $\frac{E_i^*}{E_i} = \frac{1}{1}$ | | | w = 2 mm, $\frac{E_i^*}{E_i} = \frac{1}{10}$ | | |
|---|---|---|---|---|---|---|
|  | df = 0.0125 | df = 0.025 | df = 0.05 | df = 0.0125 | df = 0.025 | df = 0.05 |
| 1$^{st}$ | 8390.45 | 8390.45 | 8390.45 | 7856.75 | 7856.75 | 7856.75 |
| f$_{max}$ [Hz] | 8289.2 | 8170.7 | 7952.5 | 7764.3 | 7650.2 | 7449.0 |
| A$_{max}$ [mm] | 0.1391 | 0.0705 | 0.0363 | 0.1515 | 0.0769 | 0.0396 |
| $p = \frac{f}{A}$ | 59611 | 115880 | 219273 | 51258 | 99491 | 188326 |
| $p^*$ | 1.1630 | 1.1647 | 1.1643 | 1.0 | 1.0 | 1.0 |
| $p_r^*$ | 98.9 | 99.8 | 99.6 | 10 | 10 | 10 |

**Table 3:** Results of peak values of frequency and displacement and their p and p$^*$ values. *df* stands for the modal damping factor.

The normalized p* values are shown to be insensitive to the value of the damping factor of the SSDM. The healing process starts at p* = 1.0 and ends at p* = 1.165. Next, the range of $1 \leq p^* \leq 1.165$ is linearly mapped into the range $10 \leq p_r^* \leq 100$, the Normalized Healing Factor (NHF), using the transformation: $p_r^* = a \cdot p^* + b$ ; $a = 545.45$ ; $b = -535.45$. A bone which reaches values above 99 can be considered as healed.

### 3.4 SSDM results

The steady state analyses were conducted for all the cases: w = 0, 0.1, 0.5, 1.0 and 2.0 mm as well, as $E_i^* = E_i$, $E_i/2$, $E_i/5$ and $E_i/10$ and the corresponding densities. 10 eigenfrequencies were used with a damping factor of 0.05. The frequencies and amplitudes at the maximum response are summarized in Table A4 in the Appendix.

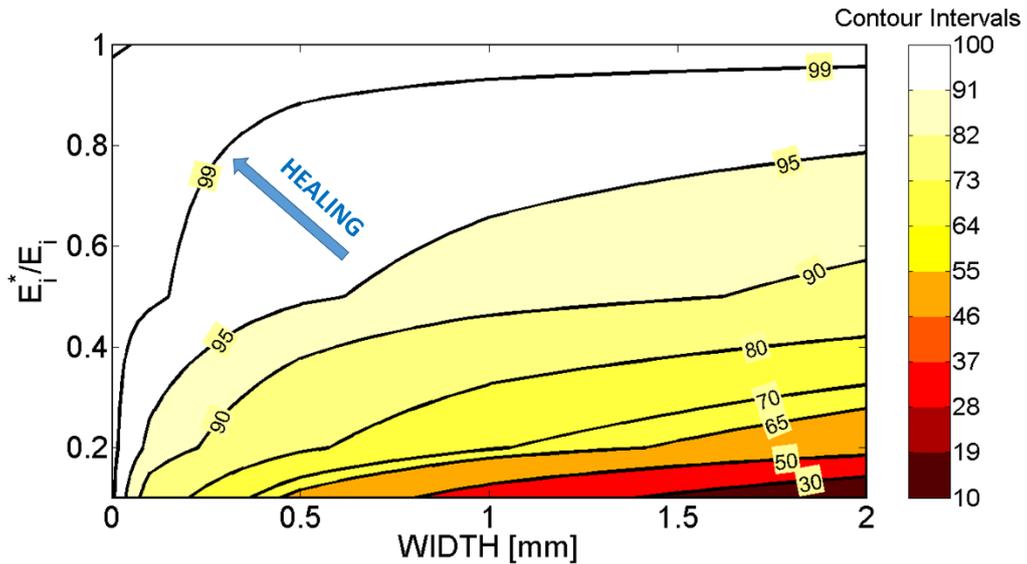

**Figure 9:** Contour maps of the NHF: $p_r^* = a \cdot p^* + b$ where a= 545.45 and b=-535.45.

Figure 9 indicates that important gradients of the parameter $p_r^*$ are obtained up to around pr*=90. Still, for this value, $E_i^*/E_i \leq 0.6$, which corresponds to the early stages of bone healing in the present study. It can be concluded that the use of the pr* (NHF) parameter might help monitoring the evolution of the bone healing process, with an improved initial resolution, while high values near 100 might indicate full healing.

## 4. DISCUSSION and CONCLUSIONS

This study provides additional results on the "classical" resonant frequency analysis with regards to dental implants stability. The approach adopted here differs from previous approaches in the sense that the stiffness and the thickness of the bone-implant interfacial region are *varied simultaneously* as opposed to an interface of vanishing thickness or conversely of a lack of well-defined interface. We examined the presence of a variable width peri-implant layer whose elastic stiffness can be degraded with respect to the pristine bone, in an attempt to model bone evolutions during its reconstruction. Our approach consists in assessing the effect of changes of bone properties in a region of interest located around the implant surface because it has been shown that the implant success depends on bone properties around the implant (at a distance of around 200 µm) and that the properties of bone tissue located farther away from the implant surface are less important regarding the implant success [11][15-17].

In this paper, we did not determine the ISQ as an indicator of primary stability in the absence of a clear mathematical definition of this quantity, in spite of its huge popularity. Instead, we characterized both the resonant frequencies and steady state dynamic analysis of the bone- implant system, since those are definitely not separate entities.

One of the first outcomes of this work is that for the system at hand, the first and the second resonant frequencies, while being inferior to 10 KHz, considered as an upper limit in the RFA systems, are in fact close to each other. It is only from the third resonance and beyond that significantly higher and distinct resonant frequency values are obtained. With that, one must keep in mind that the common RFA analysis considers essentially the first resonance, whose variations are deemed to be related to the implant stability.

Since implant stability is related to the nature of the peri-implant layer, i.e. its thickness and stiffness, one can notice from Fig. 7a that for a significantly weakened layer, there is some gradient in the first resonant frequency, however, once the bone's stiffness exceeds some 0.4 times its maximum value, the gradient fades out and the change in resonant frequency is very small, perhaps undetectable or more probably lying in the standard error range of the measurements (which is around +- 2 ISQ (Vayron et al. 2018a b)). Stated otherwise, the only identifiable layer has to be relatively wide and the peri-implant bone significantly weak. Such an observation corresponds to that of Zanetti et al. (Zanetti et al. 2018) who studied different configurations of implant geometries and weakened layers. Here, one should keep in mind that the clinically relevant peri-implant layers are of the order of 200 μm thick with a reduced stiffness of 80% that of mature bone tissue (Mathieu et al. 2011b, 2012; Vayron et al. 2018c). For such a case, the discriminative capability of the resonant frequency analysis is unlikely to provide reliable information, at least for the first 2 modes. It might be that the sought after information could be retrieved from the examination of higher resonance modes, but this has not been done yet and may prove to be experimentally cumbersome.

Let us turn now our attention to the implant displacements, corresponding here to an applied load of 1N, this result being scalable to any applied load due to the problem's linearity. Those displacements are the complementary missing side of the resonant frequency method, as implemented in commercial devices, even if such displacement values have a definite clinical relevance to bone remodeling. The numerical simulations show that the absolute values of the displacements are quite small, and therefore difficult to measure practically. The whole range of those displacements can be vastly magnified and made to vary between two arbitrary limits, 10 and 100, in a way that is probably quite similar to what is done with the ISQ. However, the thus defined "Normalized Healing Factor (NIF)", is not found to reveal more information regarding the implant stability than that obtained from the resonant frequencies. Figure 9 clearly shows that the discernable cases for which there is a clear gradient in NIF values are whose of a wide and weak peri-implant layer. Once the bone has exceeded roughly 0.4 of its original stiffness, the gradient in NIH becomes quite dull.

This observation and its matching one for resonant frequency values indicates that whatever way frequencies of displacements are reduced into a "figure of merit", the latter cannot be linearly correlated to the former, and the system is more sensitive at the very initial stages of the implantation.

The following conclusions can thus be drawn from the present study:

The sensitivity of the RFA to changes of mechanical properties of periprosthetic bone tissue seems relatively weak.

Another indicator considering the amplitude as well as the resonance frequency might be more adapted to bone healing estimations.

However, these results need to be verified experimentally as well as clinically.